\documentclass[12pt]{article}
\usepackage{amsmath}
\usepackage{amssymb}

\usepackage{graphicx}
\usepackage{mathrsfs}

\usepackage[numbers,compress]{natbib}
\begin{document}
\title{Quantum theory of redshift in de Sitter expanding universe}
\author{Ion I. Cot\u aescu\\ {\small \it  West 
                 University of Timi\c soara,}\\
   {\small \it V. P\^ arvan Ave. 4, RO-1900 Timi\c soara, Romania}}

\maketitle

\begin{abstract}
The quantum theory of the Maxwell free field in Coulomb gauge on the de Sitter expanding universe is completed with the technical elements needed for building a coherent quantum theory of redshift. Paying a special attention to the conserved observables and defining the projection operator selecting the detected momenta it is shown that the expectation values of the energies of the emitted and detected photons comply with the Lema\^ itre  rule of Hubble's law. Moreover, the quantum corrections to the dispersions of the principal observables and new uncertainty relations are derived.

Pacs: 04.62.+v
\end{abstract}


\newpage

\section{Introduction}

An important  source of empirical data in the observational astrophysics is the light emitted by different cosmic objects whose redshifts  encapsulate information about the cosmic expansion and possible peculiar velocity of the observed objects \cite{H0}.  For understanding these two contributions one combined so far the Lema\^ itre rule \cite{L1,L2} of Hubble's law \cite{Hubb}, describing the cosmological effect  \cite{H}, with the usual theory of the Doppler effect of special relativity. Recently we proposed an improvement of this approach replacing the special relativity with our de Sitter relativity \cite{CdSR1,CdSR2}. We obtained thus a rdshift formula having a new term combining the cosmological and kinetic contributions in a non-trivial manner \cite{Dop1}. Moreover, we related the black hole shadow and redshift for the Schwarzschild \cite{Dop2}  and Reissner-Nordstrom \cite{Dop3} black holes moving freely  in the de Sitter expanding universe.

The next step might be the quantum theory of redshift but this was never considered because of the real or presumed difficulties in constructing the quantum theory of light in curved backgrounds. In fact there is nothing much in it since we have already the classical and quantum theory of the free Maxwell field on the de Sitter expanding universe \cite{Max} including the de Sitter QED in the first order of perturbations \cite{CQED}. Therefore, we may build a quantum theory of redshift exploiting  this framework and solving the specific difficulties of this problem. We devote this paper to this goal constructing step by step the redshift theory from the classical level  up to a new quantum approach able to reveal the quantum corrections and the uncertainty relations related to this effect.  

The cornerstone here is the conformal covariance of the Maxwell equations in Coulomb gauge allowing us to take over the all the results of special relativity in the co-moving local charts (called here frames) with conformal coordinates of the de Sitter expanding universe \cite{Max}. In these frames the quantisation of the Maxwell field can be done in canonical manner as in special relativity. The difference is that there is a richer algebra of isometry generators  giving rise to more conserved quantities of the classical theory that become conserved one-particle operators after quantization \cite{Max}.  Of a special interest is the energy operator which does not commute with the components of the conserved momentum generating new uncertainty relations \cite{CGRG}. 

On the other hand, the conformal coordinates are different from the physical ones which are of the Painlev\' e type \cite{Pan} being related to the conformal ones through  coordinate transformations depending on time. However, in the quantum theory these   transformations change the time evolution picture as we have shown  in Refs. \cite{T1,T2,T3}. Therefore, for avoiding this difficulty,  we restrict ourselves to the conformal coordinates setting the initial conditions at the time  $t_0$ when the scale factor $a(t_0)=1$ and the physical and conformal space coordinates coincide. Under such circumstances the physical effects may be studied by using exclusively the conserved one-particle operators.

In addition, we pay attention to a pair of sensitive technical problems which are crucial in our approach. The first one is related to the momentum dependent phase of the plane wave solutions of the Maxwell equations which determines the form of the energy operator. Here we set for the first time the phase which guarantees the correct flat limit of our theory. The second problem is related to the detector measuring the redshift  which has to select only the radiation emitted by a remote source. For doing so we assume that the detector filters the momenta in a desired domain of the momentum space whose associated projection operator helps us to derive the expectation values and dispersions of the measured observables.  

We obtain thus a complete quantum theory of the redshift observed in the radiation emitted by a remote source without peculiar velocity.  We show that the expectations values of the energies of the emitted and detected photons comply with the  Lema\^ itre rule of Hubble's law while the dispersions get new quantum corrections involved in a set of new uncertainty relations. 
However, it is less probable to identify such  corrections in the astrophysical observations since these are very small in our actual expanding universe.  Nevertheless,  the methods developed here are important as these can be adapted to any spatially flat Friedman-Lema\^ itre-Robertson-Walker (FLRW) space-time including those studied in the cosmology of early universe. 

We start in the second section with a brief review of the de Sitter geometry defining the conserved quantities and introducing the conformal and physical coordinates. In the next section we revisit the  classical theory of redshift pointing out the role of the conserved quantities in deriving the Lema\^ itre equation. The fourth section is devoted to the classical theory of the Maxwell field showing how by fixing a convenient phase we assure the correct flat limit and deriving the principal conserved quantities which  become one-particle operators after the quantization performed in the next section.  In the last part of this section we show how the wave packets can be measured by choosing a suitable projection operator for selecting the momenta of the modes which contribute to the expectation values of the principal conserved observables.   The next section is devoted to the quantum redshift for which we derive the new quantum corrections and uncertainty relations. Finally we present some concluding remarks.

As here we develop a quantum approach,  we introduce a special notation denoting by $\omega_H=\sqrt{\frac{\Lambda}{3}}c$ the de Sitter Hubble constant (frequency) since $H$ is reserved for the energy or Hamiltonian operator \cite{CGRG}. Moreover,  the Hubble time  $t_H=\frac{1}{\omega_H}$ and the Hubble length   $l_H=\frac{c}{\omega_H}$ will have the same form  in the natural Planck units with $c=\hbar=G=1$ we use here.

\section{de Sitter expanding universe}

The manifold in which we would like to study the quantum theory of redshift is the expanding portion $M_+$ of the de Sitter space-time $M$ known as the de Sitter expanding universe. The manifold $M$ may be defined as the hyperboloid of radius $1/\omega_H$  in the five-dimensional flat spacetime $(M^5,\eta^5)$ of coordinates $z^A$  (labelled by the indices $A,\,B,...= 0,1,2,3,4$) having the pseudo-Euclidean metric $\eta^5={\rm diag}(1,-1,-1,-1,-1)$. The frames $\{x\}$  of coordinates $x^{\mu}$ (of natural indices $\alpha,\mu,\nu,...=0,1,2,3$) can be introduced on $M$ or $M_+$ giving the set of functions $z^A(x)$ which solve the hyperboloid equation,
\begin{equation}\label{hip}
\eta^5_{AB}z^A(x) z^B(x)=-\frac{1}{\omega_H^2}\,,
\end{equation}
where  $\omega_H$ is the Hubble de Sitter constant (frequency)  in our notations.  

In what follows we consider the co-moving frames with two sets of local coordinates,  the {\em conformal} pseudo-Euclidean ones, $\{t_c,{\bf x}_c\}$, and the physical de Sitter-Painlev\' e coordinates, $\{t,{\bf x}\}$. The conformal time $t_c$ and the conformal Cartesian spaces coordinates $x_c^i$ ($i,j,k,...=1,2,3$) are defined by the functions 
\begin{eqnarray}
z^0(x_c)&=&-\frac{1}{2\omega_H^2 t_c}\left[1-\omega_H^2({t_c}^2 - {\bf x}_c^2)\right]\,,
\nonumber\\
z^i(x_c)&=&-\frac{1}{\omega_H t}x_c^i \,, \label{Zx}\\
z^4(x_c)&=&-\frac{1}{2\omega_H^2 t_c}\left[1+\omega_H^2({t_c}^2 - {\bf x}_c^2)\right]\,,
\nonumber
\end{eqnarray}
written with the vector notation, ${\bf x}_c=(x^1_c,x^2_c,x^3_c)\in {\Bbb R}^3\subset M^5$. These frames cover the expanding portion $M_+$ for $t_c \in (-\infty,0)$ and ${\bf x}_c\in {\Bbb R}^3$ while the collapsing part $M_-$ is covered by
similar charts with $t_c >0$. In both these cases we have the same conformal flat line element,
\begin{eqnarray}
ds^{2}&=&\eta^5_{AB}dz^A(x_c)dz^B(x_c)\nonumber\\
&=&g_{\mu\nu}(x_c)\,dx_c^{\mu}dx_c^{\nu}=\frac{1}{\omega_H^2 {t_c}^2}\left({dt_c}^{2}-d{\bf x}_c\cdot d{\bf x}_c\right)\,.\label{mconf}
\end{eqnarray}
Here we restrict ourselves to the expanding portion $M_+$ which is a plausible  model of our expanding universe. 

The  de Sitter-Painlev\' e coordinates $\{t, {\bf x}\}$ on the expanding portion  can be introduced  directly by substituting
\begin{equation}\label{EdS}
t_c=-\frac{1}{\omega_H}e^{-\omega_H t}\,, \quad {\bf x}_c={\bf x} e^{-\omega_H t}\,,
\end{equation}
where $t\in(-\infty, \infty)$ is the {proper} or cosmic time while $x^i$ are the 'physical' Cartesian space coordinates. Then the line element reads
\begin{equation}\label{mdSP}
ds^2=g_{\mu\nu}(x)\,dx^{\mu}dx^{\nu}=(1-\omega_H^2 {{\bf x}}^2)\,dt^2+2\omega_H {\bf x}\cdot d{\bf x}\,dt -d{\bf x}\cdot d{\bf x}\,. 
\end{equation}
Notice that this chart is useful in applications since in the flat limit (when $\omega_H \to 0$) its coordinates become just the Cartesian ones of the Minkowski spacetime.  
In the charts with combined coordinates $\{t,{\bf x}_c\}$ the metric takes the   FLRW form 
\begin{equation}
ds^2=dt^2-a(t)^2\,d{\bf x}_c\cdot d{\bf x}_c\,, \quad a(t)=e^{\omega_H t}\,, 
\end{equation}
where $a(t)$ is the scale factor of the expanding portion which can be rewritten in the conformal chart,  
\begin{equation}\label{scale}
a(t_c)\equiv a[t(t_c)]=-\frac {1}{\omega_H t_c}\,,
\end{equation}
as a function defined for $t_c<0$.

The de Sitter space-time is a hyperbolic manifold with the maximal symmetry whose  isometry group is just the gauge group $SO(1,4)$ of the embedding manifold $M^5$  that leave  invariant its metric and implicitly Eq. (\ref{hip}). Therefore, given a system of coordinates defined by the functions $z=z(x)$, each transformation ${\frak g}\in SO(1,4)$ defines an isometry, $x\to x'=\phi_{\frak g}(x)$, derived from the system of equations
\begin{equation}\label{zz}
z[\phi_{\frak g}(x)]={\frak g}z(x)\,.
\end{equation}
The frames related through such isometries play the role of the  inertial frames similar  of special relativity. Each  isometry $x\to x'=\phi_{{\frak g}(\xi)}(x)$, depending on the group parameter $\xi$, gives rise to an associated Killing vector, ${k}=\partial_{\xi}\phi_{\xi}|_{\xi=0}$. In a canonical parametrization of the $SO(1,4)$ group with real skew-symmetric parameters $\xi^{AB}=-\xi^{BA}$, any infinitesimal isometry, 
\begin{equation}
\phi^{\mu}_{{\frak g}(\xi)}(x)=x^{\mu}+ \xi^{AB}k^{\mu}_{(AB)}(x)+....
\end{equation}
depend on the components 
\begin{equation}\label{KIL}
k_{(AB)\,\mu}= z_A\partial_{\mu}z_B-z_B\partial_{\mu}z_A\,, \quad z_A=\eta_{AB}z^B\,,
\end{equation}
of the Killing vectors associated to the parameters $\xi_{AB}$.

The classical conserved quantities along geodesics have the general form  ${\cal K}_{(AB)}(x,{\bf P})=\omega_H k_{(AB)\,\mu} p^{\mu}$ where the four-momentum components  $p^{\mu}=\frac{dx^{\mu}(s)}{d\lambda}$ are the derivatives with respect to the afine parameter $\lambda$ which satisfies $ds=m d\lambda$ such that   $g_{\mu\nu}p^{\mu}p^{\nu}=m^2$. The conserved quantities with physical meaning \cite{CGRG} are, the energy $E=\omega_H  k_{(04)\,\mu}p^{\mu}$, the angular momentum components,  $L_i= \frac{1}{2}\,\varepsilon_{ijk}k_{(jk)\,\mu}p^{\mu}$, and the components $K_i=  k_{(0i)\,\mu} p^{\mu}$ and $R_i=  k_{(i4)\,\mu} p^{\mu}$ of two vectors  related to  the conserved momentum ${\bf P}$ and its associated dual momentum ${\bf Q}$ as,  \cite{CGRG}
\begin{equation}\label{PQKR}
{\bf P}=-\omega_H ({\bf R}+{\bf K})\,, \quad {\bf Q}=\omega_H ({\bf K}-{\bf R})\,. 
\end{equation}
satisfying the identity
\begin{equation}\label{disp}
E^2-\omega_H^2 {{\bf L}}^2-{\bf P}\cdot {\bf Q}=m^2\,,
\end{equation}
corresponding to the first Casimir invariant of the $so(1,4)$ algebra \cite{CGRG}. In the flat limit, $\omega_H\to 0$  and $-\omega_H t_c\to 1$, we have ${\bf Q} \to {\bf P}$ such that this identity  becomes just  the usual null mass-shell condition $E^2-{\bf P}^2=m^2$ of  special relativity. The conserved quantities ${\cal K}_{(AB)}$ transform as a five-dimensional skew-symmetric tensor  under the $SO(1,4)$  transformations generating the de Sitter isometries. 

The simple isometries  we meet here are  the translations of parameters ${\bf d}=(d^1,d^2,d^3)$ that act as
\begin{equation}\label{trans}
\begin{array}{lll}
t_c&=&t_c'\\
{\bf x}_c&=&{\bf x}_c'+{\bf d}
\end{array} \to
\begin{array}{lll}
t&=&t'\\
{\bf x}&=&{\bf x}'+{\bf d}\,e^{\omega_H t}
\end{array} \,.
\end{equation}
Hereby we can derive the Killing vector of components
\begin{equation}\label{KIK1}
k_j^0=0\,, ~~~k_j^i=\left.\frac{\partial x_c^i}{\partial d^j}\right|_{{\bf d}=0}=\delta^i_j=\omega_H\left(k^i_{(0,j)}+k^i_{(i,4)}\right)\,,
\end{equation}
where $k^i_{(AB)}=g^{ij}(x_c)\,k_{(AB)\,j}$ result from  Eq. (\ref{KIL}). This  give rise to the conserved momentum of the classical approach and to the momentum operator of the quantum theory. We shall see in the next section  that these isometries  transform the energy, angular momentum and dual momentum but preserving the conserved momentum. 

\section{Null geodesics and redshift}

We consider now  the null geodesics of the photons (with $m=0$) denoting the conserved quantities along these geodesics with capital letters. In the conformal chart the four-momentum components, denoted now by  $k^{\mu}_c=\frac{dx^{\mu}_c}{d\lambda}$, satisfy the identity
\begin{equation}\label{int1}
g_{\mu\nu}(x_c)\,k_c^{\mu}k_c^{\nu}=a(t_c)^2\left[k_c^0(t_c)^2-{\bf k}_c(t_c)^2\right]=0\,,
\end{equation}
resulted from the line element (\ref{mconf}). In addition, we may consider the   components 
\begin{equation}\label{Pci}
P^i=-g_{jk}(x_c) k_i^j \frac{dx_c^k}{d\lambda}= a(t_c)^2\frac{dx_c^j}{d\lambda}\,,
\end{equation} 
of the {\em conserved} momentum ${\bf P}={\bf n}_P P$ ($P=|{\bf P}\,|$).  Then by using the prime integrals (\ref{int1}) and (\ref{Pci}) we derive the {\em energy} and {\em covariant} momentum  in this frame as
\begin{eqnarray}
k_c^0(t_c)&=&\frac{dt_c}{d\lambda}=\frac{P}{a(t_c)^2} =\omega_H^2t_c^2 P\,,\label{uu0}\\
{\bf k}_c(t_c)&=&\frac{d{\bf x}_c}{d\lambda}= \frac{\bf P}{a(t_c)^2}=\omega_H^2t_c^2{\bf  P}\,.\label{uui}
\end{eqnarray}
The null geodesic results simply  as \cite{CdSG}
\begin{equation}\label{null1}
{\bf x}_c(t_c)={\bf x}_{c0}+{\bf n}_P(t_c-t_{c0})\,,
\end{equation}
concluding that this geodesic is determined completely by the unit vector ${\bf n}_P$   and the initial condition ${\bf x}_c(t_{c0})={\bf x}_{c0}$.

The corresponding physical quantities measured in the chart  $\{t,{\bf x}\}$, may be obtained by substituting the physical coordinates according to Eq. (\ref{EdS}). Thus we find 
\begin{eqnarray}
k^0(t)&=&\frac{dt}{d\lambda}=Pe^{-\omega_H t}\,,\label{uu01}\\
{\bf k}(t)&=&\frac{d{\bf x}}{d\lambda}= {\bf P}\,e^{-\omega_H t} +\omega_H {\bf x}(t)\,Pe^{-\omega_H t}\,,\label{uui1}
\end{eqnarray}
which represent the measured energy and covariant momentum in the point $[t, {\vec x}(t)]$ of the null geodesic \cite{CdSG}
\begin{equation}
{\bf x}(t)={\bf x}_0 e^{\omega_H(t-t_0)}+{\bf n}_P\,\frac{e^{\omega_H(t-t_0)}-1 }{\omega_H}\,,\label{geodE}
\end{equation}
which is passing through the space point ${\bf x}(t_0)={\bf x}_0$ at the initial time $t_0$. The conserved quantities on this geodesic can calculated at any time as
\begin{eqnarray}
E&=&e^{-\omega_H t}\left[P+\omega_H {\bf x}(t)\cdot {\bf P} \right]\,,\label{Ene1}\\
{\bf L}&=&{\bf x}(t)\land {\bf P}e^{-\omega_H t}\,,\label{L1}\\
{\bf Q}&=&2\omega_H {\bf x}(t)E+{\bf P} e^{-2\omega_H t}[1-\omega_H^2{\bf x}(t)^2]\,,\label{Q1}
\end{eqnarray}
observing that these satisfy the identity (\ref{disp}) for $m=0$.

The momentum defined by Eq. (\ref{uui1}) can be split  as  ${\bf k}(t)=\hat{\bf k}(t)+ \bar{\bf k}(t)$ where
\begin{equation}\label{ppx}
\hat {\bf k}(t)={\bf P} e^{-\omega_H t}\,,\quad \bar{\bf k}(t)=\omega_H\,{\bf x}(t) \, k^0(t)\,,
\end{equation}
are the {\em peculiar} and respectively {\em recessional} momenta we have defined recently \cite{Kin}. The prime integral derived from the line element (\ref{mdSP}) gives the familiar identity
\begin{equation}\label{Pan1}
k^0(t)^2- \hat{\bf k}(t)^2=0\,, 
\end{equation}
which is just the mass-shell condition of special relativity satisfied by the energy and peculiar momentum along the null geodesics. 

Now we may analyse how two different observers  measure  a photon moving on a null geodesic  which is passing through the origins $O$ and $O'$ of their proper co-moving frames $\{t,{\bf x}\}_{O}$ and $\{t,{\bf x}'\}_{O'}$.  We assume that the photon is emitted in $O'$  at the initial time $t_0$ when the origin  $O'$ is translated with respect to $O$ as  
\begin{equation}\label{tr}
{\bf x}(t_0)={\bf x}'(t_0)+{\bf d}\,e^{\omega_H t_0}\,, 
\end{equation}
where the translation parameter ${\bf d}={\bf n}\, d$ of the isometry (\ref{trans}) has the direction $OO'$ given by the unit vector  ${\bf n}$. If we know that the photon is emitted in ${\bf x}'(t_0)=0$ with the  momentum ${\bf k}=-{\bf n}\,k$ and energy $k^0=k$  we ask  which are the energy and momentum of this photon  measured in the origin $O$  at the final time $t_f$ when the particle reach this point.  For solving this problem we look first for the conserved momentum that is the same in the points  $O'$ and $O$, 
\begin{equation}\label{Pn}
{\bf P}'={\bf P}={\bf k}\,e^{\omega_M t_0}~ ~\to~~P=k\,e^{\omega_M t_0} \,,~~~{\bf n}_P=-{\bf n}\,,
\end{equation}
since this is invariant under translations being  associated to their generators. 
Furthermore, we observe that the choice of the initial time
\begin{equation}\label{tin}
t_{c0}=-\frac{1}{\omega_H} ~~\to~~ t_0=0\,,
\end{equation}
when $a=1$ and, consequently,  the conformal and physical space coordinates {\em coincide}. This simplifies the calculations allowing us to find the quantities measured by the observers $O$ and $O'$  at this moment derived from Eqs. (\ref{Ene1})-(\ref{Q1}).  

The results are presented in the next table where we introduce intuitive notations for the initial, $E_i$, and final, $E_f$, photon energies which are the physical quantities involved in redshift.
\begin{center}
\begin{tabular}{lll}
&frame $\{t,{\bf x}'\}_{O'}$ &frame $\{t,{\bf x}\}_{O}$\\
&&\\
initial condition &${\bf x}'(0)=0$&${\bf x}(0)={\bf d}$\\
conserved energy& $E'\equiv E_i=k$&$E\equiv E_f=k(1-\omega_H d)$\\
conserved momentum &${\bf P}'={\bf k}$&${\bf P}={\bf k}$\\
angular momentum&${\bf L}'=0$&${\bf L}=0$\\
dual momentum&${\bf Q}'={\bf k}$&${\bf Q}={\bf k}(1-\omega_H d)^2$
\end{tabular}
\end{center}
We observe that the translation (\ref{tr}) change the energy and the components of the adjoint momentum but preserving the invariant (\ref{disp}) with $m=0$. Indeed, in the frame $\{t,{\bf x}'\}_{O'}$ we have $E_i^2={\bf k}^2={\bf P}'\cdot{\bf Q}'$. 
In the frame $\{t,{\bf x}\}_{O}$ the photon  arrives in ${\bf x}=0$ at the  conformal time $t_{cf}=d-\frac{1}{\omega_H}$ corresponding to the cosmic time
\begin{equation}
t_f=-\frac{1}{\omega_H}\ln(-\omega_H t_{cf})=-\frac{1}{\omega_H}\ln (1-\omega_H d)\,.
\end{equation}
Hereby we can deduce that $O$ measures the final peculiar momentum,
\begin{equation}
\hat {\bf k}(t_f)={\bf k}(1-\omega_H d)\,,
\end{equation}
resulted from Eqs. (\ref{ppx}) for $t=t_f$. Obviously, this satisfies the mass-shell condition $E_f^2=\hat {\bf k}(t_f)^2={\bf P}\cdot {\bf Q}$. Hereby we conclude that  ${\bf P}$ and ${\bf Q}$ are conserved quantities that cannot be measured directly but  
complete each other for closing the identity (\ref{disp}). The only measurable quantities remain thus the energy, peculiar momentum and angular momentum.

These results allow us to recover the Lema\^ itre expression of Hubble's law giving the redshift $z$ as
\begin{equation}\label{LH}
\frac{1}{1+z}=\frac{E_{f}}{E_i}=1-\omega_H d=1-\frac{d}{l_H}\,,
\end{equation}
and  the physical distance observer-source at the time $t_f$, 
\begin{equation}
d_f=\frac{d}{-\omega_H t_{cf}}=\frac{d}{1-\omega_H d}\,, 
\end{equation}
which was increasing because of  the space expansion during the photon propagation.
Thus we revisited the redshift in the particular case when $O'$ does not have a peculiar velocity. The general result for arbitrary peculiar velocity of $O'$  was derived recently  \cite{Dop1,Dop2}. 

These results that hold at the level of the geometric optics in de Sitter background are incomplete since at this level we neglect the wave behaviour and polarization that may be studied considering  the classical and quantum theory of the Maxwell field.

\section{Classical Maxwell field}

Let us consider first the classical approach  denoting by $A$ the electromagnetic potential of the Maxwell field minimally coupled to the de Sitter gravity, whose action in an arbitrary chart $\{x\}$ of $(M,g)$ reads
\begin{equation}\label{action}
\mathcal{S}[A]=\int d^{4}x \sqrt{g}\,{\cal L}=-\frac{1}{4}\int d^{4}x
\sqrt{g}\,F_{\mu \nu  }F^{\mu \nu  } ,
\end{equation}
where $g=|\det(g_{\mu\nu})|$ and $F_{\mu \nu  }=\partial_{\mu } A_{\nu
}-\partial_{\nu } A_{\mu }$ is the field strength. From this action one derives
the field equations
\begin{equation}
\partial_{\nu}(\sqrt{g}\,g^{\nu\alpha}g^{\mu\beta}F_{\alpha\beta})=0\,,
\end{equation}
which are invariant under conformal transformations, $g_{\mu\nu}\to
g_{\mu\nu}'=\Omega g_{\mu\nu}$ and 
\begin{equation}\label{conf}
A_{\mu}\to A'_{\mu}=A_{\mu}\quad A^{\mu}\to A^{\prime\,\mu}=\Omega^{-1} A^{\mu}\,.
\end{equation}
The canonical variables $A_{\mu}$ must obey, in addition, the
Lorentz condition
\begin{equation}\label{Lor}
\partial_{\mu}(\sqrt{g}\,g^{\mu\nu}A_{\nu})=0\,,
\end{equation}
which is no longer conformally invariant since
\begin{equation}\label{Lor1}
\partial_{\mu}(\sqrt{g'}\,g^{\prime\,\mu\nu}A_{\nu}')=\partial_{\mu}(\sqrt{g}\,g^{\mu\nu}A_{\nu})+
\sqrt{g}A^{\mu}\partial_\mu\Omega\,.
\end{equation}
However, we may get over this inconvenience imposing the Coulomb gauge, $A_0=0$,  in the conformal chart $\{t_{c},\bf{x}_c\}$ since then the second term of Eq. (\ref{Lor1})  does not contribute. 

\subsection{Plane waves}

Under such circumstances we can write the solutions of the Maxwell equations in Coulomb gauge, $(\partial_{t_c}^2-\Delta_c)A_i=0$, and Lorentz condition, $\partial_{x_{c}^i} A_i=0$, taking over all the well-known results of special relativity. Thus we may write the plane wave solutions 
\begin{eqnarray}
&&A_i(x_c)=A_i^{(+)}(x_c)+A_i^{(-)}(x_c)\nonumber\\
&&=\int d^3k \sum_{\lambda}\left[e_i({\bf n}_k,\lambda) \hat f_{\bf k}(x_c) \hat a({\bf
k},\lambda)+[e_i({\bf n}_k,\lambda)\hat f_{\bf k}(x_c)]^* \hat a^*({\bf
k},\lambda)\right]\,,~~~~\label{field1}
\end{eqnarray}
in terms of wave functions in momentum representation, $\hat a({\bf k},\lambda)$,
polarization vectors, $e_i({\bf n}_k,\lambda)$, and fundamental solutions of
the d'Alambert equation,
\begin{equation}\label{fk}
\hat f_{\bf k}(x_c)=\frac{1}{(2\pi)^{3/2}}\frac{1}{\sqrt{2k}}\,e^{i\delta({\bf k})-ikt_c+i{\bf k}\cdot
{\bf x}_c}\,,
\end{equation}
where ${\bf k}=k {\bf n}_k$ is the momentum vector with $k=|{\bf k}|$. 

The momentum-dependent phase $\delta({\bf k})$ is introduced  in order to assure the correct flat limit of the plane wave solutions when  $\omega_H\to 0$.  According to Eqs. (\ref{EdS}), we see that the entire phase of the function (\ref{fk}) behaves as
\begin{equation}
i\delta({\bf k})-i k \left(-\frac{1}{\omega_H}+t \right)+i {\bf x}_c\cdot {\bf k} + {\cal O} (\omega_H)\,, 
\end{equation}
having a pole in $\omega_H=0$. For removing this singularity we must impose the condition  
\begin{equation}\label{phase}
\lim_{\omega_H\to 0}\left(\delta({\bf k})+\frac{k}{\omega_H}\right)=0\,,
\end{equation}
giving the correct flat limit of special relativity, 
\begin{equation}
\lim_{\omega_H\to 0}e^{i\delta({\bf k})-ikt_c+i{\bf k}\cdot
{\bf x}_c}=e^{-ikt+i{\bf k}\cdot {\bf x}}\,.
\end{equation}
For avoiding some difficulties related to this explicit  phase, it is convenient to redefine
\begin{eqnarray}
f_{\bf k}(x_c)&=&e^{-i\delta({\bf k})}\hat f_{\bf k}(x_c)=\frac{1}{(2\pi)^{3/2}}\frac{1}{\sqrt{2k}}\,e^{-ikt_c+i{\bf k}\cdot {\bf x}_c}\,,\label{fk1} \\
a({\bf k},\lambda)&=&e^{i\delta({\bf k})}\hat a({\bf k},\lambda)\,,\label{aa} 
\end{eqnarray}
substituting $\hat f_{\bf k}(x_c)\hat a({\bf k},\lambda)=f_{\bf k}(x_c) a({\bf k},\lambda)$  in Eq. (\ref{field1}). In Ref. \cite{Max} we neglected the problem of this phase which is solved here for the first time. This guarantees the correct flat limit as in the case of the rest frame vacua we defined recently for the massive Dirac \cite{V1}, Klein-Gordon \cite{V2} and Proca \cite{V3} showing that only these vacua assure correct flat limits \cite{V4}.

The functions  $f_{\bf k}(x)$  are assumed to be of positive frequencies while
those of negative frequencies are $f_{\bf k}(x)^*$. These solutions satisfy the
orthonormalization relations \cite{Max}
\begin{eqnarray}
\left(  f_{\bf k},f_{{\bf k}'}\right)=-\,\left(  f_{\bf k}^*,f_{{\bf
k}'}^*\right)&=&\delta^3({\bf k}-{\bf k}')\,,\\
\left( f_{\bf k},f_{{\bf k}'}^*\right)&=&0\,,
\end{eqnarray}
and the completeness condition
\begin{equation}\label{comp}
i\int d^3k\,  f^*_{\bf k}(t_c,{\bf x}_c)
\stackrel{\leftrightarrow\,\,}{\partial_{t_c}} f_{\bf k}(t_c,{\bf
x}')=\delta^3({\bf x}_c-{\bf x}_c')\,,
\end{equation}
with respect to the Hermitian form 
\begin{equation}\label{fg}
\left(f,g\right)=i\int d^3x_c\, f^*(t_c,{\bf x}_c) \stackrel{\leftrightarrow\,\,\,}{\partial_{t_c}} g(t_c,{\bf x}_c)\,.
\end{equation} 
where we denote $f\stackrel{\leftrightarrow}{\partial} g=f\partial g-g \partial f$.

The polarization vectors ${\bf e}({\bf n}_k,\lambda)$ in Coulomb gauge must be
orthogonal to the momentum direction,
\begin{equation}
{\bf k}\cdot{\bf e}({\bf n}_k,\lambda)=0\,,
\end{equation}
for any polarization $\lambda=\pm 1$. We remind the reader that the
polarization can be defined in different manners independent of the
form of the scalar solutions $f_{\bf k}$. In general, the
polarization vectors have c-number components which must satisfy
\cite{rf:19}
\begin{eqnarray}
{\bf e}({\bf n}_k,\lambda)\cdot{\bf e}({\bf
n}_k,\lambda')^*&=&\delta_{\lambda\lambda'}\,,\\
 \sum_{\lambda}e_i({\bf n}_k,\lambda)\,e_j({\bf
n}_k,\lambda)^*&=&\delta_{ij}-\frac{k^i k^j}{k^2}\,.\label{tran}
\end{eqnarray}
Here we restrict ourselves to consider only the {\em circular} polarization for
which the supplementary condition ${\bf e}({\bf n}_k,\lambda)^*\land {\bf
e}({\bf n}_k,\lambda)=i \lambda\, {\bf n}_k$ is requested.

We obtained thus mode expansions in terms of transverse plane waves of given momentum and helicity. The functions $w_{i({\bf k},\lambda)}=e_i({\bf n}_k,\lambda)f_{\bf k}$, are of positive frequencies while those of negative frequencies
are  $w_{i({\bf k},\lambda)}^*$.  We say that these sets of fundamental
solutions of the Maxwell equation define the {\em momentum-helicity} basis. Note that an energy-helicity basis can also be defined \cite{Max}.

\subsection{Conserved quantities}

The Maxwell theory is invariant under $SO(1,4)$  isometries such that under a given isometry , $x\to x'=\phi_{{\frak g}(\xi)}(x)$, depending on the group parameter $\xi$, 
the vector field $A$ transforms as $A\to A' =T_{\xi}A$,  according to
the operator-valued representation $\xi\to T_{\xi}$ of the isometry
group defined by the well-known rule
\begin{equation}
\frac{\partial \phi^{\nu}_{\xi}(x)}{\partial
x_{\mu}}\left(T_{\xi}A\right)_{\nu}[\phi(x)]=A_{\mu}(x)\,.
\end{equation}
The corresponding generator, $X_{K}=i\,\partial_{\xi}T_{\xi}|_{\xi=0}$, has the
action
\begin{equation}\label{XA}
(X_{k}\, A)_{\mu}=-i({k}^{\nu}A_{\mu;\nu}+{k}^{\nu}_{~;\mu}A_{\nu})\,.
\end{equation}
where $k$ is the Killing vector associated to $\xi$. In a canonical parametrization of the $SO(1,4)$ group  we have the correspondence
\begin{equation}
\xi^{AB}~~\to~~k_{(AB)}~~\to~~X_{k_{(AB)}}\equiv X_{(AB)}\,,
\end{equation}
which means that the generators $X_{(AB)}$ form a basis of the vector representation of the $so(4,1)$ algebra carried by the space of the vector potential, $A$.
This algebra yields the principal observables, i. e.  the energy
operator $\hat H=\omega X_{(05)}$, the components of the momentum, 
$\hat P^i=-\omega_H(X_{(i4)}+X_{(0i)})$,  dual momentum,  $\hat Q^i=\omega_H(X_{(0i)}-X_{(i4)})$, and  total angular momentum, $\hat J_i=\frac{1}{2}\varepsilon_{ijk}X_{(jk)}$ 
($i,j,...=1,2,3$), operators \cite{rf:7,CGRG}. 

The action of these operators can be calculated according to Eq. (\ref{XA}) using the concrete form of the Killing vectors derived from Eq. (\ref{KIL}). In what follows we restrict ourselves to present the results in the Coulomb gauge (with $A_0=0$) which are useful in applications. The energy and momentum operators do not have spin parts, acting as \cite{Max}
\begin{eqnarray}
\hat H\,  A_{j}&=&-i\omega_H\left(t_c \frac{\partial}{\partial{t_c}} + x_c^i\frac{\partial}{\partial x_c^i}+1\right) { A}_{j}\,,\label{HH}\\
\hat P^i  A_{j}&=&-i\frac{\partial}{\partial x_c^i}\, A_{j}\,,\label{PP}
\end{eqnarray}
while the action of the total angular momentum reads
\begin{equation}
\hat J_i\,  A_j=\hat L_i  A_j
-i\varepsilon_{ijk} A_k\,,\label{J}
\end{equation}
where $\hat {\bf L}={\bf x}_c\times {\bf P}$ is the usual angular momentum operator.
In addition, we define the Pauli-Lubanski (or helicity) operator $\hat W=\hat{\bf
P}\cdot \hat{\bf J}$ whose action depends only on the spin parts,
\begin{equation}
\hat W  A_i=\varepsilon_{ijk}\frac{\partial}{\partial x_c^j}  A_k\,.
\end{equation}
This operator will define the  polarization in the canonical basis of the $so(3)$ algebra as in special relativity \cite{Max}. 

However, the dual momentum has the most complicated action,
\begin{eqnarray}
\hat Q^i A_j &=&-i\omega_H^2 \left[ 2x_c^i \left(t_c\frac{\partial}{\partial t_c}+x_c^k\frac{\partial}{\partial x_c^k} \right)+(t_c^2-{\bf x}_c^2)\frac{\partial}{\partial x_c^i}\right]A_j \nonumber\\ 
&&-i \omega_H^2\left(2\delta _{ij}{\bf x}_c\cdot {\bf A} +x_c^i A_j-x_c^j A_i \right)\,,
\end{eqnarray}
we present here for the first time.

We constructed thus the basis $\{\hat H, \hat P^i,\hat Q^i,\hat J_i\}$ of the vector representation of the $so(1,4)$ algebra whose commutation rules are
\begin{eqnarray}
&&\left[\hat H, \hat J_i \right]=0\\
&&\left[ \hat H, \hat P^i \right]=i\omega_H  \hat P^i\,,\quad \hspace*{8.5 mm} \left[
 \hat H,  \hat Q^i \right]=-i\omega_H \hat Q^i\,,\label{HPQ}\\
&&\left[\hat J_i ,\hat  P^j \right]=i\varepsilon_{ijk}\hat P^k\,,\quad~~~~~
\left[ \hat J_i, \hat Q^j \right]=i\varepsilon_{ijk} \hat Q^k\,,\label{PJP}\\
&&\left[\hat Q^i,\hat  Q^j \right]=0\,,\quad \hspace*{17mm} \left[\hat P^i,\hat P^j\right]=0\,,\\
&&\left[\hat Q^i, \hat P^j \right]=2 i \omega_H \delta_{ij}\hat H + 2 i
\omega_H^2 \varepsilon_{ijk} \hat J_k \,,
\end{eqnarray}
from which we deduce 
\begin{equation}
\left[\hat P^i, \hat W \right]=0 \,, \quad \left[\hat H, \hat W \right]= i \omega_H \hat W\,,
\end{equation}
understanding that the maximal set of commuting operators we may use is $\{\hat P^i, \hat W\}$. The first Casimir operator of this algebra has the form \cite{CGRG}
\begin{equation}
{\cal C}_1=\hat H^2+3 i \omega_H \hat H-\hat{\bf Q}\cdot\hat{\bf P}
-\omega_H^2\hat {\bf J}\cdot\hat{\bf J}
\end{equation}
giving the supplemental equation ${\cal C}_1 A_i=0$ which in the Coulomb gauge is just the d'Alambert one. 

In what concerns the structure of the $so(1,4)$ algebra we observe that there are two Abelian sub-algebras generated by the momentum components, $\{\hat P^i\}$, and respectively by those of the dual momentum, $\{\hat Q^i\}$. Another specific feature is that the energy operators does not commute with the momentum components as in special relativity. Regarding our notations we must specify that the upper or lower  positions of the space indices do not have here a meaning but we denoted the components of the momentum and dual momentum with upper indices since in the flat limit these become $\hat Q^i\to \hat P^i\to -i\partial_i$,  i. e. the contravariant space components of the momentum operator with respect to the Minkowski metric.   

The conserved quantities of our Lagrangian theory are related to the operators of the $so(1,4)$ algebra via Noether's theorem.  From the action (\ref{action}) we deduce that in Coulomb's gauge the conserved quantities can be derived by using the Hermitian form (\ref{fg}) as  \cite{Max}
\begin{equation}\label{Ck1}
X~~\to~~C[X]=\frac{1}{2}\, \delta_{ij} \left( A_i,X
A_j\right)  \quad \forall X \in so(1,4)\,,
\end{equation}
This integral  can be expressed  in terms of  electric and  magnetic  components of the field strength \cite{CQED} or as a mode integral in momentum representation. Thus we can conclude that we outlined here a coherent de Sitter electrodynamics which has a correct flat limit.

\section{Quantum Maxwell field}

The next step is the quantization we may perform in canonical manner as in special relativity exploiting the global conformal invariance of the theory in Coulomb gauge.
There are many delicate problems that can be avoided if we restrict ourselves to the conserved operators in the Heisenberg picture assuming that the  quantum states are defined at the initial time (\ref{tin}) when $a(0)=1$ and the conserved and peculiar momenta coincide. Then we have to focus only on the conserved operators calculated at this moment which are enough for analysing  the quantum redshift.   However,   this method is not suitable for studying other dynamic operators as, for example, the coordinate operator and that of peculiar momentum. 

\subsection{One-particle operators}

We assume that the wave functions $a$ of the field (\ref{field1})  becomes field operators (with $a^{*}\to a^{\dagger}$) \cite{rf:19} such that the potentials (\ref{field1}) become field operators denoted by ${\cal A}_i$. We assume that the field operators fulfill the standard commutation relations in the momentum-helicity basis from which the non-vanishing ones are
\begin{equation}\label{com1}
[a({\bf k},\lambda), a^{\dagger}({\bf k}^{\,\prime},\lambda ')]
=\delta_{\lambda\lambda '} \delta^3 ({\bf k}-{\bf k}^{\,\prime})\,.
\end{equation}
Then the  Hermitian field ${\cal A}={\cal A}^{\dagger}$ is correctly quantized according to the {\em canonical} rule
\begin{equation}
[ {\cal A}_i(t_c,{\bf x}_c),\pi^j(t_c,{\bf x}_c')]=[ {\cal A}_i(t_c,{\bf
x}),\partial_{t_c}{\cal A}_j(t_c,{\bf x}_c')]=i\,\delta^{tr}_{ij}({\bf x}_c-{\bf x}_c')\,,
\end{equation}
where
\begin{equation}
\pi^j=\sqrt{g}\,\frac{\delta {\cal L}}{\delta
(\partial_{t_c}{\cal A}_j)}=\partial_{t_c}{\cal A}_j
\end{equation}
is the momentum density in Coulomb gauge (${\cal A}_0=0$) and
\begin{equation}
\delta^{tr}_{ij}({\bf x}_c)=\frac{1}{(2\pi)^3}\int d^3q
\left(\delta_{ij}-\frac{q^iq^j}{q^2}\right)e^{i {\bf q}\cdot {\bf x}_c}
\end{equation}
is the well-known transverse $\delta$-function \cite{rf:19} arising
from Eq. (\ref{tran}).

As in special relativity we consider a unique  vacuum state, $|0\rangle$, of the Fock such that 
\begin{equation}
a({\bf k},\lambda)\,|0\rangle=0\,,\quad \langle 0|\,a^{\dagger}({\bf
k},\lambda)=0\,.
\end{equation}
The sectors with a given number of particles may be constructed using the standard methods for obtaining the generalized momentum-helicity basis of the the Fock space. 

The one-particle operators corresponding to the conserved quantities
(\ref{Ck1}) can be calculated in Coulomb gauge as \cite{Max}
\begin{equation}\label{opo}
{\cal X}=\frac{1}{2}\,\delta_{ij}:\left( {\cal A}_i, X {\cal A}_j\right)
\end{equation}
respecting the normal ordering of the operator products \cite{rf:19}. The obvious
algebraic properties
\begin{eqnarray}
[{\cal X}, {\cal A}_i(x)]&=&-X {\cal A}_i(x)\,, \\
\left[{\cal X}, {\cal Y}\,\right]&=&\frac{1}{2}\,\delta_{ij}:\left( {\cal A}_i, [X,Y]{\cal A}_j\,\right)\label{algXX}
\end{eqnarray}
are due to the canonical quantization adopted here. However, there are many
other conserved operators which do not have corresponding differential
operators but can be defined directly as mode expansions.  The simplest example is the operator of the number of particles,
\begin{equation}
{\cal N}=\int d^3k \sum_{\lambda}   a^{\dagger}({\bf k},\lambda) a({\bf
k},\lambda)\,,
\end{equation}

The conserved one-particle operators whose mode expansions can be derived easily are the components of the conserved momentum operator,
\begin{equation}
{\cal P}^l=\frac{1}{2}\,\delta_{ij}:\left({\cal A}_i,  \hat P^l {\cal A}_j\right)\,=\int d^3k\,
k^l \sum_{\lambda}  a^{\dagger}({\bf k},\lambda) a({\bf k},\lambda)\,,
\end{equation}
and the Pauli-Lubanski operator,
\begin{equation}
{\cal W}=\frac{1}{2}\,\delta_{ij}:\left( {\cal A}_i, \hat W {\cal A}_j\right)\,=\int d^3k\,k
 \sum_{\lambda}\lambda\,  a^{\dagger}({\bf k},\lambda) a({\bf k},\lambda)\,,
\end{equation}
which commutes with the momentum componenta, $[{\cal P}^i, {\cal W}]=0$. The  momentum-helicity basis
\begin{equation}\label{basis}
|0\rangle,\, |{\bf k},\lambda\rangle=a^{\dagger}({\bf k},\lambda)|0\rangle,\, 
 |{\bf k},\lambda;{\bf k}',\lambda'\rangle=a^{\dagger}({\bf k},\lambda) a^{\dagger}({\bf k}',\lambda')|0\rangle,\,...
\end{equation}
 is formed by the eigenvectors of the set of commuting operators  $\{ {\cal W},\,{\cal P}^i\}$ corresponding to the discrete polarizations, $\lambda,\, \lambda+\lambda',\, ...\in {\Bbb Z}$  and momenta ${\bf k},\, {\bf k}+{\bf k}',\,...$ of continuous spectrum ${\Bbb R}^3_k$.

The problem of the energy operator is more delicate but can be solved resorting to the identity 
\begin{equation}\label{Hf}
(\hat H f_{\bf k})(x)=-i\omega_H \left(k^i\partial_{k_i}+{\frac{3}{2}}\right)f_{\bf k}(x) \,,
\end{equation}
satisfied by the functions (\ref{fk1}).  Then, after a few manipulation and applying the Green theorem, we obtain the final result \cite{Max},
\begin{eqnarray}
{\cal H}&=&\frac{1}{2}\,\delta_{ij}:\left( {\cal A}_i, \hat H {\cal A}_j\right)\,: \nonumber\\
&=&\frac{i\omega_H}{2}\int d^3k\, k^i  \sum_{\lambda}\,
a^{\dagger}({\bf
k},\lambda)\stackrel{\leftrightarrow}{\partial}_{k_i} a({\bf
k},\lambda)\,.\label{Hpp}
\end{eqnarray}
Hereby we see that the form of the energy operator is strongly dependent on the phase  of the operators  $a({\bf k},\lambda)$ defined by Eq. (\ref{aa}). This is in accordance with the similar property of the energy operator that holds for the  Klein-Gordon, Dirac or Proca free fields on this background. This general behaviour is due to the space expansion giving the  dependence of the energy operator on the translations that change the phase. Nevertheless, this behaviour does not change the commutation relations 
\begin{eqnarray}
\left[{\cal H},{\cal P}^i \right]&=& i\,\omega_H {\cal P}^i\,,\label{Hppc}\\
\left[{\cal H},{\cal W}\right]&=& i\,\omega_H {\cal W}\,,\label{Wpp}
\end{eqnarray}
which are independent on the phase $\delta({\bf k})$ as it results from Eqs.  (\ref{HPQ}) and (\ref{algXX}).

\subsection{Wave packets and measurements}

A simple model which prevents us from complicated calculations is that of the one-particle wave-packets. In our Heisenberg picture these are given by the  time-independent  one-particle states, 
\begin{equation}
|\alpha\rangle = \int d^3 k \sum_{\lambda}\alpha_{\lambda}({\bf k})\, a^{\dagger}({\bf k},\lambda) |0\rangle
\end{equation}
defined by the square integrable functions in momentum representation $\alpha_{\lambda}({\bf k})$ which must satisfy the normalization condition
\begin{equation}
\langle\alpha|\alpha\rangle= \int d^3 k \sum_{\lambda}|\alpha_{\lambda}({\bf k})|^2=1\,.
\end{equation}  
The corresponding 'wave functions' 
\begin{equation}
A[\alpha]_i(x)=\langle 0|{\cal A}_i(x)|\alpha\rangle=\int d^3k \sum_{\lambda}e_i({\bf n}_k,\lambda) f_{\bf k}(x) \alpha_{\lambda}({\bf k})\,,
\end{equation}
are known as wave-packets. These are useful auxiliary functions related to those of the momentum representation through the inversion relations  
\begin{equation}
\alpha_{\lambda}({\bf k})=\delta_{ij}e_i({\bf n}_k,\lambda)^*\left( f_{\bf k}, A[\alpha]_j\right)\,.
\end{equation}
Moreover, the expectation values of the one-particle operators (\ref{opo}) in the state $|\alpha\rangle$ can be calculated simply as
\begin{equation}\label{simple}
\langle \alpha|{\cal X}|\alpha\rangle = \delta_{ij}\left( A[\alpha]_i, XA[\alpha]_j\right) \,,
\end{equation}
avoiding the tedious algebra of field operators. 

Once the wave-packet is prepared this evolves causally until an ideal apparatus measures some parameters. More specific, this  apparatus can measure all the eigenvalues  of the operators ${\cal W}$ and  ${\cal P}^i$ which are diagonal in the momentum-helicity basis.   In an experiment we can set this apparatus to select only the momenta included in a desired domain $\Delta \subset {\Bbb R}^3_k$  by using a suitable projection operator $\Lambda_{\Delta}=\Lambda_{\Delta}^{\dagger}$   (satisfying $\Lambda_{\Delta}^2=\Lambda_{\Delta}$) that  can be represented as
\begin{equation}
\Lambda_{\Delta}=|0\rangle\langle 0|+\int_{\Delta} d^3k \sum_{\lambda} a^{\dagger}({\bf k},\lambda)\,|0\rangle\langle 0|\, a({\bf k},\lambda)+... \,,
\end{equation}
where the integral is restricted to the domain $\Delta$. During the experiment this operator  filters only the momenta ${\bf k}\in\Delta$ transforming the state of the system as $|\alpha\rangle \to \Lambda_{\Delta}|\alpha\rangle$. Then the expectation value $\langle{\cal X}\rangle$ of a one-particle operator ${\cal X}$ can be calculated as \cite{Mes}
\begin{equation}\label{expv}
\langle{\cal X}\rangle=\frac{\langle \alpha| \Lambda_{\Delta} {\cal X}|\alpha\rangle}{\langle \alpha| \Lambda_{\Delta}|\alpha\rangle}
\end{equation}
taking into account that our one-particle operators commute with $\Lambda_{\Delta}$. The quantity
\begin{equation}
\langle \alpha| \Lambda_{\Delta}|\alpha\rangle= \int_{\Delta} d^3 k \sum_{\lambda}|\alpha_{\lambda}({\bf k})|^2\le 1\,,
\end{equation}
gives the probability $P_{\Delta}=| \langle \alpha| \Lambda_{\Delta}|\alpha\rangle |^2$ of measuring any momentum ${\bf k}\in \Delta$. 
Obviously, when we can measure the whole continuous spectrum, $\Delta={\Bbb R}^3_k$, then $\Lambda_{\Delta}\to{\bf 1}$, $P_{\Delta}=1$  and $\langle{\cal X}\rangle=\langle \alpha| {\cal X}|\alpha\rangle$.  

Furthermore, bearing in mind the role of the phase factor in Eq. (\ref{aa}) we assume that the functions $\alpha_{\lambda}$ have the general form
\begin{equation}\label{aak}
\alpha_{\lambda}({\bf k})=e^{i\delta({\bf k})}\hat\alpha_{\lambda}({\bf k}) \,, 
\end{equation} 
where $\hat\alpha_{\lambda}=\hat\alpha_{\lambda}^*$ are real valued functions. Then we can derive the expectation values of the operators which are diagonal in the basis (\ref{basis}) by using the rule (\ref{simple}) as 
\begin{eqnarray}
\langle ({\cal P}^i)^n\rangle&=&\frac{1}{\langle \alpha| \Lambda_{\Delta}|\alpha\rangle} \int_{\Delta} d^3 k  (k^i)^n\sum_{\lambda}\hat\alpha_{\lambda}({\bf k})^2\,,\label{Pn}\\
\langle {\cal W}^n\rangle&=&\frac{1}{\langle \alpha| \Lambda_{\Delta}|\alpha\rangle} \int_{\Delta} d^3 k  \sum_{\lambda}  \lambda^n \hat\alpha_{\lambda}({\bf k})^2\,,\label{Wn}
\end{eqnarray} 
For the energy operator which is not diagonal in this basis we may apply the same formula but using, in addition, the identity (\ref{Hf}) and the Green theorem which helps us to write  
\begin{eqnarray}
\langle {\cal H}\rangle & =&\frac{1}{\langle \alpha| \Lambda_{\Delta}|\alpha\rangle}\left\{\frac{i\omega_H}{2}\int_{\Delta} d^3k\, k^i  \sum_{\lambda}\,
\alpha^{*}_{\lambda}({\bf k})\stackrel{\leftrightarrow}{\partial}_{k_i} \alpha_{\lambda}({\bf k})\right\}\nonumber\\
&=&-\frac{1}{\langle \alpha| \Lambda_{\Delta}|\alpha\rangle}\left\{\omega_H \int_{\Delta} d^3k\left[ k^i\partial_{k^i}\delta({\bf k})\right] \sum_{\lambda}\,\hat \alpha_{\lambda}({\bf k})^2\right\} \,,\label{Hpp3}
\end{eqnarray}
since $\hat \alpha_{\lambda}^*\stackrel{\leftrightarrow}{\partial}\hat \alpha_{\lambda}=0$ as these are  real valued functions. Note that the operator $k^i\partial_{k^i}$ in momentum space is in fact a radial operator such that this does not affect the polarization vectors which depend only on the unit vector of the momentum direction, ${\bf n}_k$. Finally,  by using again  Eq. (\ref{simple}) we may write  
\begin{equation}
\langle\alpha|{\cal H}^2|\alpha\rangle =\delta_{ij}\left(\hat H A[\alpha]_i, \hat HA[\alpha]_j\right) \,,
\end{equation}
which helps us to obtain the useful formula 
\begin{eqnarray}
\langle {\cal H}^2\rangle & =&\frac{\omega_H^2}{\langle \alpha| \Lambda_{\Delta}|\alpha\rangle}\left\{\int_{\Delta} d^3k\left[ k^i\partial_{k^i}\delta({\bf k})\right]^2 \sum_{\lambda}\,
\hat \alpha_{\lambda}({\bf k})^2 \right.\nonumber\\
&&\hspace*{21mm}+\left.\int_{\Delta} d^3k\,   \sum_{\lambda}\,
\left[\left(k^i{\partial}_{k^i}+\frac{3}{2}\right) \hat\alpha_{\lambda}({\bf k})\right]^2\right\}\,,\label{Hpp4}
\end{eqnarray}  
we need in the next application. 

\section{Quantum redshift}

Let us come back now to the problem of two translated observers, $O'$ and $O$, preparing and respectively measuring a photon state $|\alpha\rangle$.   We assume that these observers use the same global ideal apparatus represented by the operator algebra $\mathbb{A}'\cup \mathbb{A}$ formed by two sub-algebras including the field operators and the conserved ones for which we use the self-explanatory notations
\begin{eqnarray}
O' :&\quad {\cal A}'(x_c'),...{\cal H}', {\cal P}^{\prime\,i}, {\cal Q}^{\prime\,i}, {\cal L}^{\prime}_i...&\in \mathbb{A}'\,, \\
O :&\quad {\cal A}(x_c),...{\cal H}, {\cal P}^{i}, {\cal Q}^{i}, {\cal L}_i....&\in \mathbb{A}\,, 
\end{eqnarray}
The state $|\alpha\rangle$ is prepared at the initial time (\ref{tin}) when the observers are translated each other with the position vector ${\bf d}$. The translation generators ${\cal P}^{\prime\,i}={\cal P}^{i}$, which are the same in both the above sub-algebras, define the translation operator 
\begin{equation}
{\cal T}({\bf d})=\exp\left(i d^i{\cal P}^i\right)\,,
\end{equation} 
which transforms these sub-algebras,  ${\cal T}({\bf d}) :\, \mathbb{A}' \to \mathbb{A}$, such that any operator ${\cal X}'\in  \mathbb{A}'$ is transformed into ${\cal X}=T({\bf d}){\cal X}'T({\bf d})^{\dagger}\in  \mathbb{A}$. Particularly, the energy operator is translated as 
\begin{equation}\label{OpH}
{\cal H}={\cal T}({\bf d}) {\cal H}' {\cal T}({\bf d})^{\dagger}={\cal H}'+\omega_H d^i {\cal P}^i\,,
\end{equation}
according to the commutation rule (\ref{Hppc}). In what follows,  we simplify the geometry be choosing an orthogonal frame 
$\{{\bf e}_1,{\bf e}_2,{\bf e}_3\}$ such that ${\bf d}=d {\bf e}_3$. 

\subsection{Expectation values}

The observer $O'$ prepares the state $|\alpha \rangle$  in his proper co-moving frame where the principal parameters are the expectation values of  energy, $E'$, momentum components,  $P^{\prime\,i}$, and polarization, $W'$. These quantities can be calculated by using the simple rule (\ref{simple}) taking into account that the packet is defined by the functions (\ref{aak}) whose phase must be fixed according to the condition (\ref{phase}). The simplest choice is 
\begin{equation}\label{ph}
\delta({\bf k})=-\frac{k}{\omega_H}\,,
\end{equation}
since then the expectation value of the energy reads
\begin{equation}
E'\equiv \langle\alpha| {\cal H}'|\alpha\rangle=\delta_{ij}\left( A[\alpha]_i, \hat H A[\alpha]_j\right)=\int d^3k \,k \sum_{\lambda} \hat\alpha_{\lambda}({\bf k})^2
\end{equation}
as it results from Eq. (\ref{Hpp3}) for $\Delta=\mathbb{R}^3_k$. The other expectation values can be derived simpler by using Eq. (\ref{simple}) as 
\begin{eqnarray}
P^{\prime\,i}&\equiv&\langle\alpha| {\cal P}^i|\alpha\rangle=\delta_{ij}\left( A[\alpha]_i, \hat P^i A[\alpha]_j\right)=\int d^3k \,k^i \sum_{\lambda} \hat\alpha_{\lambda}({\bf k})^2\,,~~~~\\
W^{\prime}&\equiv&\langle\alpha| {\cal W}^i|\alpha\rangle=\delta_{ij}\left( A[\alpha]_i, \hat W A[\alpha]_j\right)=\int d^3k \sum_{\lambda}\lambda \,\hat\alpha_{\lambda}({\bf k})^2\,,
\end{eqnarray}
since these do not depend on the phase (\ref{ph}). We observe that in our framework with the phase (\ref{ph}) all these expectation values have the same forms as in Minkowski space-time.

However, these quantities are not accessible to the observer $O$ which focuses on the observables, ${\cal H}'$, ${\cal H}$, ${\cal P}^i$ and ${\cal W}$ selecting only the photons coming from the source $O'$, whose momenta are parallel with ${\bf e}_3$. This means that the domain of  momenta measured by $O$ is
\begin{equation}\label{Del}
\Delta=\left\{{\bf k}\, \left\vert\, -\frac{\Delta k}{2}\le k^1\le \frac{\Delta k}{2},\right.\,-\frac{\Delta k}{2}\le k^2\le \frac{\Delta k}{2},\,k^3<0\right\}
\end{equation} 
where $\Delta k$ is a small quantity. Then we may evaluate the  integrals over $\Delta$ as 
\begin{eqnarray}
\int_{\Delta}d^3k F({\bf k})&=& \int_{-\frac{\Delta k}{2}}^{\frac{\Delta k}{2}}dk^1  \int_{-\frac{\Delta k}{2}}^{\frac{\Delta k}{2}}dk^2\int_{-\infty}^{0} dk^3 F(k^1,k^2,k^3)\nonumber \\
& \simeq&(\Delta k)^2 \int_{0}^{\infty} dk F(0,0,-k)\,,
\end{eqnarray} 
according to the mean value theorem. 

Now we come back to our intuitive notations of Sec. 3 of the expectation values of the initial,  $E_i\equiv\langle {\cal H'}\rangle$, and final, $E_f\equiv\langle {\cal H}\rangle$,  energies related to the conserved momentum  of components  
$P^i\equiv\langle {\cal P}^i\rangle$ that can be observed by $O$. These expectation values have to be calculated according to Eq. (\ref{expv}) with the state  $|\alpha\rangle$ defined by the functions  (\ref{aak}) with the phase (\ref{ph}) and the projection operator $\Lambda_{\Delta}$ corresponding to the domain (\ref{Del}). First  we find that 
\begin{equation}
\langle \alpha| \Lambda_{\Delta}|\alpha\rangle= \int_{\Delta} d^3 k \sum_{\lambda}\hat\alpha_{\lambda}({\bf k})^2=(\Delta k)^2 \kappa
\,,
\end{equation}
where
\begin{equation}
\kappa= \int_{0}^{\infty} dk \sum_{\lambda} \hat\alpha_{\lambda}(0,0,-k)^2\,.
\end{equation}
Furthermore, we calculate  the expectation values defined by Eq. (\ref{Pn}) for $n=1$  
\begin{eqnarray}
P^3\equiv\langle {\cal P}^3\rangle&=&-\frac{1}{\kappa}\int_{0}^{\infty} dk\,k \sum_{\lambda} \hat\alpha_{\lambda}(0,0,-k)^2 \,,\\
\langle {\cal P}^1\rangle&=&\langle {\cal P}^2\rangle =0\,,\label{P23}
\end{eqnarray}
which do not depend on the phase (\ref{ph}). For the energy operators the situation is different since their expectation values depend on this phase  as in Eq. (\ref{Hpp3}) which allows us to write 
\begin{equation}\label{fin0}
E_i\equiv \langle {\cal H}'\rangle= \frac{1}{\kappa}\int_{0}^{\infty} dk\,k \sum_{\lambda} \hat\alpha_{\lambda}(0,0,-k)^2=-P^3\,,
\end{equation}
deriving the expectation value of Eq. (\ref{OpH}) as
\begin{equation}\label{fin1}
E_f\equiv\langle {\cal H}\rangle =(1-\omega_H d) E_i\,,
\end{equation}
recovering thus  the Lema\^ itre form of Hubble's law (\ref{LH}). 
Note that this result  can be derived in a different manner observing that
\begin{equation}
\langle\alpha|{\cal H}|\alpha\rangle=\langle\alpha|T({\bf d}){\cal H'}T({\bf d})^{\dagger}|\alpha\rangle=\langle\tilde\alpha|{\cal H}'|\tilde\alpha\rangle\,,
\end{equation}
where now the translated state $|\tilde\alpha\rangle$ is given by the functions (\ref{aak}) in which we must substitute 
\begin{equation}\label{phtr}
\delta({\bf k}) \to \tilde\delta({\bf k})=-\frac{k}{\omega_H}-{\bf k}\cdot {\bf d}\,.
\end{equation}
With this new phase Eq. (\ref{Hpp3}) gives just the result (\ref{fin1}). We must specify that now the initial energy $E_i$ observed by $O$ is different from $E'$ measured by $O'$ in contrast with the classical approach where these two quantities coincide (as in the table of Sec. 3). 

In our experiment we select only the momenta oriented along ${\bf e}_3$ such that  the polarizations vectors ${\bf  e}\,({\bf e}_3,\pm 1)=\frac{1}{\sqrt{2}}({\bf e}_1\mp i{\bf e}_2)$ are in the plane $\{{\bf e}_1,{\bf e}_2\}$.  The expectation value of the Pauli-Lubanski operator,
\begin{equation}
W\equiv \langle{\cal W}\rangle= \frac{1}{\kappa}\int_{0}^{\infty} dk\left[  \hat\alpha_{1}(0,0,-k)^2-\hat\alpha_{-1}(0,0,-k)^2\right]
\end{equation}
suggests us to introduce  the polarization angle $0\le \theta(k)\le \frac{\pi}{2}$ such that 
\begin{equation}
\hat\alpha_{1}(0,0,k)=\cos\theta(k)\, \alpha(k)\,,\quad  \hat\alpha_{-1}(0,0,k)=\sin\theta(k)\, \alpha(k)
\end{equation}
where the new function $\alpha(k)$ satisfies $ \int_{0}^{\infty} dk \, \alpha (k)^2=\kappa$. In the particular case when $\theta$ is a constant independent on $k$  we have $W=\cos 2\theta$.

\subsection{Dispersions and uncertainty}

The next step is to study the dispersions of the observables measured by $O$  applying the well-known rule
\begin{equation}
{\rm disp} {\cal X}=\left (\Delta {\cal X}\right)^2=\langle {\cal X}^2\rangle- \langle {\cal X}\rangle^2\,.
\end{equation}
We observe first that the operators ${\cal H}'$ and ${\cal H}$ commute alike with ${\cal P}^i$ and ${\cal W}$ as in Eqs. (\ref{Hppc}) and (\ref{Wpp}) but do not commute with each other since
\begin{equation}
\left[{\cal H}', {\cal H}\right]=i\omega_H^2 d^i {\cal P}^i=i\omega_H^2 d\,{\cal P}^3\,.
\end{equation}
Therefore, from the above equation and  Eq. (\ref{Hppc})  we obtain  the uncertainty relations 
\begin{eqnarray}
{\rm disp} {\cal H}'\, {\rm disp} {\cal P}^i&\ge&\frac{1}{4}\,\omega_H^2 \left| \langle {\cal P}^i\rangle \right|^2\,,\label{H0P}\\
{\rm disp} {\cal H}\, {\rm disp} {\cal P}^i&\ge&\frac{1}{4}\,\omega_H^2 \left| \langle {\cal P}^i\rangle \right|^2\,,\label{H1P}\\
{\rm disp} {\cal H}'\, {\rm disp} {\cal H}&\ge&\frac{1}{4}\,\omega_H^4 d^2 \left| \langle {\cal P}^3\rangle \right|^2\,,\label{H01P}
\end{eqnarray}
For $i=1,2$ Eqs. (\ref{P23}) allow us to set ${\rm disp} {\cal P}^1={\rm disp} {\cal P}^2=0$ without violating the 
uncertainty relations but along the third axis the relation (\ref{H0P})  is non-trivial since $P\equiv|\langle {\cal P}^3\rangle| \not=0$. On the other hand, from Eq. (\ref{Hpp4}) with the phase (\ref{ph}) we obtain 
\begin{equation}\label{aux}
\langle {{\cal H}'}^2\rangle=\langle ({\cal P}^3)^2\rangle +\omega_H^2 \chi\,.
\end{equation}
The last terms of the above equations represents the quantum corrections which are proportional with the dimensionless quantity
\begin{equation}\label{chi}
\chi=\frac{1}{\kappa}\int_{0}^{\infty} d k\,   \sum_{\lambda}\,
\left[\left(k{\partial}_{k}+\frac{3}{2}\right) \hat\alpha_{\lambda}(0,0,-k)\right]^2\,,
\end{equation}
resulted from the last term of Eq.  (\ref{Hpp4}). This is generated by the de Sitter gravity and depends exclusively on the form of the functions $\hat\alpha_{\lambda}$. 
For deriving the dispersion of the operator ${\cal H}$ we calculate first the expectation value (\ref{Hpp4}) with the new phase (\ref{phtr}) obtaining the identity 
\begin{equation}\label{aux1} 
\langle {\cal H}^2\rangle=(1-\omega_H d)^2\langle ({\cal P}^3)^2\rangle +\omega_H^2 \chi\,.
\end{equation}
Finally, from Eqs. (\ref{aux}) and (\ref{aux1})  combined with Eqs. (\ref{fin0}) and respectively (\ref{fin1}) we find  
\begin{eqnarray}
{\rm disp} E_i&\equiv & {\rm disp}{\cal H}'={\rm disp} P+\omega_H^2 \chi\,,\label{Fin}\\
{\rm disp}E_f&\equiv & {\rm disp}{\cal H}=(1-\omega_H d)^2 {\rm disp}P+\omega_H^2 \chi\,,\label{Fin1}
\end{eqnarray}
where we denote ${\rm disp}P\equiv {\rm disp}{\cal P}^3$.

Now we come back to the uncertainty relations (\ref{H0P}) and (\ref{H1P}) by using  Eqs. (\ref{Fin}) and (\ref{Fin1})  for deriving the inequalities
\begin{eqnarray}
{\rm disp} E_{i,f}\left({\rm disp} E_{i,f}-\omega_H^2\chi\right)&\ge&\frac{1}{4}\,\omega_H^2 E_{i,f}^2\,,\label{H0P1}\\
{\rm disp} P\, \left({\rm disp} P+\omega_H^2\chi\right)&\ge&\frac{1}{4}\,\omega_H^2  P^2\,,\label{H1P1}
\end{eqnarray}
from which we deduce 
\begin{eqnarray}
{\rm disp} E_{i,f}&\ge& \frac{\omega_H}{2}\left(\sqrt{E_{i,f}^2+{\omega_H^2} \chi^2}+{\omega_H} \chi\right)\nonumber\\
&=&\frac{\omega_H}{2}\left( E_{i,f}+\omega_H\chi + \frac{1}{2E_{i,f}}\,\omega_H^2\chi^2\right)+{\cal O}(\omega_H^4\chi^4)\,,\\
{\rm disp} P&\ge& \frac{\omega_H}{2}\left(\sqrt{P^2+{\omega_H^2} \chi^2}-{\omega_H} \chi\right)\nonumber\\
&=&\frac{\omega_H}{2}\left( P-\omega_H\chi + \frac{1}{2P}\,\omega_H^2\chi^2\right)+{\cal O}(\omega_H^4\chi^4)\,,
\end{eqnarray}
relating thus the dispersions to the corresponding  expectation values. More interesting is Eq. (\ref{H01P}) as depending explicitly on the distance $d$ between $O$ and $O'$. This can be rewritten  in our new notations as, 
\begin{equation}
{\rm disp} E_i\, {\rm disp} E_f\ge\frac{1}{4}\,\omega_H^4 d^2 P^2\,,\label{H01P1}
\end{equation}
and can be seen as the starting point for deriving new inequalities depending on $d$  by using Eqs. (\ref{Fin}) and (\ref{Fin1}). Our preliminary calculations indicate that these  are more complicated requiring a special analytical an numerical study which will be performed elsewhere.   

Finally, let us analyse the dispersion  of the Pauli-Lubanski operator in the simple case when the polarization angle $\theta$ is independent on $k$. We have seen that then  the expectation value has the form  $W\equiv\langle {\cal W} \rangle =\cos 2\theta$. Moreover, from Eqs. (\ref{Wn}) we obtain 
\begin{equation}
\langle {\cal W}^2 \rangle =1~~\to~~ {\rm disp} {\cal W}=\sin^2 2\theta\,,
\end{equation}
while  from Eq. (\ref{Wpp}) we derive the uncertainty  relation
\begin{equation}
{\rm disp}{\cal H}\, {\rm disp} 
{\cal W}\ge \frac{1}{4}\,\omega_H^2 |\langle {\cal W} \rangle|^2
\end{equation}
giving the restriction 
\begin{equation}\label{ttt}
\tan 2\theta\ge \frac{1}{2}\,\frac{\omega_H}{\Delta E_f}\,,  
\end{equation}
preventing one from measuring total polarizations, i. e. $\theta=0$ for $\lambda=1$ or $\theta=\frac{\pi}{2}$ for $\lambda=-1$. 
  
 \section{Concluding remarks}

We presented the complete classical and quantum theory of the Maxwell field
minimally coupled to the gravity of the de Sitter expanding universe focusing on the principal effect due to the space expansion, namely the redshift for which we derived the quantum corrections and the principal uncertainty relations.  

In the actual expanding universe the quantum  corrections and the limits of the uncertainty relations are extremely small since the actual value of $\omega_H$ (or $\hbar\,\omega_H$ in SI units) is of the order $10^{-33} eV$  such that it is less probable to  be identified in astronomical observations. Moreover,  the limitation predicted by the inequality (\ref{ttt}) in an ideal universe is too small to be separated from other polarization effects produced by the cosmic dust and plasma.  

However, these results  are interesting as coming from the first complete and coherent classical and quantum theory of the Maxwell field coupled to the gravity of an expanding universe.  The methods developed here can be applied to any spatially flat FLRW expanding universe including  the actual models of early universe.  

On the other hand, the method of regularization of the momentum dependent phase plays the same role as the rest frame vacuum of the massive particles assuring the correct flat limit of the Maxwell field. Thus we obtain a coherent quantum theory on the de Sitter expanding universe in which we may apply the perturbation methods of the traditional theory in Minkowski space-time. For this reason we hope that our approach will open the door to a large field of applications not only  in  astrophysics and cosmology but even in particle physics.

\end{document}